\documentclass[useAMS,usenatbib]{mn2e}
\usepackage{graphicx} 
\usepackage{color} 

\title{Microscopic calculation of the equation of state of nuclear
matter and neutron star structure}
\author[S. Gandolfi, 
A. Yu. Illarionov, 
S. Fantoni, 
J.C. Miller,
F. Pederiva,
K.E. Schmidt
]
{S. Gandolfi$^{1,2,3}$\thanks{E-mail: stefano@lanl.gov},
A.Yu. Illarionov$^{4,2,3}$,
S. Fantoni$^{2,3,5}$,
J.C. Miller$^{2,3,6}$,
\newauthor
F. Pederiva$^{4,7}$,
K.E. Schmidt$^{8}$
\\
$^{1}$Theoretical Division, Los Alamos National Laboratory, Los Alamos, NM 
87545, USA\\
$^{2}$International School for Advanced Studies, SISSA, Via Beirut
2/4 I-34014 Trieste, Italy \\
$^{3}$INFN, Sezione di Trieste, Trieste, Italy \\
$^{4}$Dipartimento di Fisica dell'Universit\`{a} di Trento, via Sommarive
14, I-38123 Povo, Trento, Italy \\
$^{5}$INFM {\sl DEMOCRITOS} National Simulation Center, Via
Beirut 2/4 I-34014 Trieste, Italy \\
$^{6}$Department of Physics (Astrophysics), University of Oxford, Keble 
Road, Oxford OX1 3RH, UK \\
$^{7}$INFN, Gruppo Collegato di Trento, Trento, Italy \\
$^{8}$Department of Physics, Arizona State University, Tempe, AZ 85287, 
USA \\
}
\begin{document}

\date{}

\maketitle

\begin{abstract}
 We present results for neutron star models constructed with a new 
equation of state for nuclear matter at zero temperature. The ground state 
is computed using the Auxiliary Field Diffusion Monte Carlo (AFDMC) 
technique, with nucleons interacting via a semi-phenomenological 
Hamiltonian including a realistic two-body interaction. The effect of 
many-body forces is included by means of additional density-dependent 
terms in the Hamiltonian. In this letter we compare the properties of the 
resulting neutron-star models with those obtained using other nuclear 
Hamiltonians, focusing on the relations between mass and radius, and 
between the gravitational mass and the baryon number.
 \end{abstract}


\section{Introduction}
 While real neutron stars are very complicated objects, their main global 
properties can usually be well-approximated by considering simple 
idealized models consisting of perfect fluid in hydrostatic equilibrium. 
If rotation can be neglected to a first approximation (as is the case for 
the spin rates of most currently-known pulsars) then the model can be 
taken to be spherical and its structure obtained by solving the 
Tolman-Oppenheimer-Volkoff (TOV) equations, enabling one to calculate, for 
example, the stellar mass as a function of radius or of central density. 
However, this requires specification of an equation of state (EOS) for the 
neutron-star matter which clearly plays a fundamental role in determining 
the properties of the resulting models. The EOS needs to take account of 
the strong spin-isospin correlations induced by realistic interactions, 
with particular regard to the tensor ones \citep{raffelt96}.

The EOS can in principle be computed by means of many-body theories using 
effective density-dependent interactions, such as those given by Skyrme 
forces but the phenomenological nature of these can be a disadvantage in 
making reliable calculations of neutron-star properties \citep{stone03}. 
Making a microscopic calculation based on a Hamiltonian describing the 
properties of light nuclei and symmetric nuclear matter (SNM), is both 
challenging and of great relevance. Variational approaches using 
correlated basis functions (CBF) and Fermi Hyper Netted Chain techniques 
(FHNC)  \citep{fantoni98} are good candidates for doing this but the 
strong spin-isospin dependence of the nuclear Hamiltonian requires the 
introduction of approximations into the FHNC scheme which cannot be fully 
controlled, such as the single operator chain (SOC) 
\citep{Pandharipande79}.

The Auxiliary Field Diffusion Monte Carlo method (AFDMC) \citep{schmidt99} 
has been employed for computing the ground state energy of nuclei, giving 
very good agreement with other existing accurate techniques in few-body 
physics \citep{gandolfi07b} and highlighting important limitations of 
other many-body theories used for calculations of both nuclear structure 
\citep{gandolfi07} and neutron matter \citep{gandolfi09}. In this letter 
we present results from an AFDMC calculation of the EOS of 
$\beta$-equilibrium matter (relevant for neutron stars), based on a new 
class of non-relativistic Hamiltonians using nuclear potentials with a 
realistic two-body interaction, taken from the Urbana-Argonne scheme 
\citep{wiringa95} and incorporating many-body interactions via 
density-dependent terms.

Modern nuclear Hamiltonians are usually constrained so as to reproduce 
properties of light nuclei as measured in laboratory experiments 
\citep{pieper01}, but the densities involved there are much lower than 
those found in neutron-star cores for which it becomes necessary to 
introduce a three-body potential, such as UIX or IL1-IL5 \citep{pieper01}. 
However, these give very large and very different energy contributions to 
the EOS of pure neutron matter (PNM) at high density 
\citep{sarsa03,fantoni07,gandolfi07c}. Also, recent AFDMC results by 
\citet{gandolfi07} indicate that SNM is not well reproduced by the 
FHNC/SOC techniques used to constrain these Hamiltonians, which are 
therefore not very suitable for calculating neutron-star models.

Because of this, we have proceeded here in a different way, using 
density-dependent terms to simulate many-body forces with the forms used 
coming from explicit integration of these forces over the variables of 
particle 3 for the three-body force, over those of particles 3 and 4 for 
the four-body force, and so on. In bulk matter, the neutron and proton 
densities $\rho_n$ and $\rho_p$ can be taken to be constant quantities,
whereas in confined systems, like nuclei, they are operators.

As a first step in this direction, we have revisited the {\sl old} 
three-parameter density-dependent LP model of \citet{lagaris81} and 
\citet{friedman81}, with the values of the free parameters fixed so as to 
reproduce the saturation point and compressibility of SNM. The LP
calculations were performed using the FHNC/SOC approximation and we then 
re-fitted the density-dependent term using the AFDMC calculations. By 
considering the chemical potentials, we have constructed an EOS for a 
mixture of protons, electrons and muons in $\beta$-equilibrium and have 
then used the resulting EOS for constructing neutron-star models. We 
stress that (i) the ground states of nuclear matter as a function of 
$\rho_p$ and $\rho_n$ have been calculated with the same AFDMC method 
which has been shown to be very accurate for calculating the ground states 
of various nucleonic systems, and (ii) the many-body part of the nuclear 
interaction has been self-consistently determined by these solutions.

\section{The model}
 
We model the EOS by simulating nuclear matter using a finite number of 
interacting nucleons in a periodic box. The number of particles is chosen 
from among the magic numbers giving a rotationally invariant wave function 
for the corresponding non-interacting system and the volume of the box is 
fixed by the density. For the two-body interaction, we take the Argonne 
AV6$'$ potential \citep{wiringa02}, which includes the four central 
spin-isospin components and the two tensor ones. The six components of the 
long range OPEP potential are fully included, whereas only the first six 
of the 18 components of the intermediate and short range parts of the AV18 
interaction \citep{wiringa95}, $v_I^p(r_{ij})$ and $v_S^p(r_{ij})$, are 
kept. The corresponding amplitudes $I_p$ and $S_p$ are re-fitted so as to 
correctly reproduce the deuteron properties and to give the best fit to NN 
scattering data \citep{wiringa02}.
 
The many-body interactions are represented by density-dependent factors 
of the structural form given by the LP model. The resulting potential, 
denoted as DD6$'$, is given by the following six two-body components
 \begin{eqnarray}
v_{DD6'}^p &=& v_{OPEP}^p +  v_I^p e^{-\gamma_1\rho} + v_S^p + 
{\rm TNA}(\rho) \,, \nonumber \\
{\rm TNA}(\rho) &=& 
3\gamma_2\rho^2e^{-\gamma_3\rho}\left(1-\frac{2}{3}\left(\frac{\rho_n-\rho_p}{\rho_n+\rho_p}\right)^2\right)
\end{eqnarray}
with $\gamma_1$, $\gamma_2$ and $\gamma_3$ being fixed so as to reproduce 
the experimental values of the saturation density $\rho_0=0.16$ fm$^{-3}$, 
the binding energy per particle $E_0=-16$ MeV and the compressibility 
$K=9\rho_0^2\left({\partial^2 
E(\rho)}/{\partial\rho^2}\right)_{\rho_0}\approx240$ MeV. (Note that in 
this paper we follow the nuclear-physics convention of using $\rho$ to 
refer to a particle number density rather than a mass density.)

Many nuclear matter calculations of the 1980s were made with the LP model 
interaction, and gave very good agreement with data for large nuclei, as 
discussed by \citet{benhar93} and \citet{pandharipande97}. Afterwards, 
mainly because of the difficulty of treating the density as an operator 
when dealing with nuclei, this model was forgotten and attention was 
switched to the development of increasingly sophisticated three-body 
potentials. On the basis of the results obtained with AFDMC quantum 
simulations, we believe that we should instead now proceed in the 
direction of constructing increasingly sophisticated density-dependent 
two-body effective interactions, which may also account for N-body 
interactions with $N\ge 3$.

A particular limitation of DD6$'$ is the exclusion of the non-local 
components of AV18. The most important of these is the spin-orbit term, 
particularly when dealing with nuclei. We estimate that the missing 
non-local components contribute to the EOS of neutron matter by no more 
than $5\%$ \citep{gandolfi09}. Work towards introducing the spin-orbit 
components is in progress. A second important point concerns the 
structural form of the density-dependent terms and one may want to 
increase the number of Feynman diagrams included in the construction of 
UIX and contributing to the many-body interaction.

\section{Calculation of the equation of state of nuclear matter}

We computed the ground state of the system using the AFDMC method (for 
details, see \citealp{gandolfi07c} and references therein), simulating SNM 
with $A=28$ nucleons in a periodic box, as described by 
\citet{gandolfi07}. The next magic number of nucleons providing small 
finite-size corrections is 132, as shown by \citet{gandolfi09}, which 
requires an unjustified computational effort given the restriction to 
local components in the DD6$'$ model. Table \ref{tab:DDI} gives the values 
of the free parameters of DD6$'$ corresponding to the best fit to 
experimental data for $\rho_0$, $E_0$ and $K$, and compares these with the 
values for the original LP model.

\begin{table}
\begin{center}
 \caption{The AFDMC results for the free parameters of the DD6$'$ 
interaction as compared with the original values of the LP model 
calculated within the FHNC/SOC approximation.}
 \begin{tabular}{ccc}
\hline
$parameter$  & $FHNC/SOC$ & $AFDMC$  \\
\hline
$\gamma_1$ & 0.15 fm$^3$ & 0.10 fm$^3$ \\
$\gamma_2$ & -700 fm$^6$ & -750 fm$^6$ \\
$\gamma_3$ & 13.6 fm$^3$ & 13.9 fm$^3$ \\
\hline
\label{tab:DDI}
\end{tabular}
\end{center}
\end{table}

The AFDMC density-dependent terms give more attraction than in the 
original LP model, consistent with the fact that FHNC/SOC overbinds SNM. 
The TNA term is $\sim 30\%$ larger at $\rho_0$ and more than the double at 
$5\rho_0$, giving more attraction, and $\exp(\gamma_1\rho)-1$ is $\sim 
30\%$ smaller over the whole range $(\rho_0,5\rho_0)$, giving less 
repulsion.

We find that the AFDMC results for the binding energy per nucleon of SNM 
at densities larger than $\sim 0.08$ fm$^{-3}$ can be very well described 
by:
 \begin{equation}
E_{SNM}(\rho)=E_0+a(\rho-\rho_0)^2+b(\rho-\rho_0)^3e^{\gamma(\rho-\rho_0)}
\,,
\end{equation}
where $E_0=-16.0$ MeV, $\rho_0=0.16$ fm$^{-3}$, $a=520.0$ MeV fm$^6$, 
$b=-1297.4$ MeV fm$^9$ and $\gamma=-2.213$ fm$^3$. This parametrization was 
chosen to represent the EOS of nuclear matter, reproducing properties 
constrained by terrestrial experiments on nuclei \citep{danielewicz02}.

The DD6$'$ Hamiltonian was then used to compute the EOS of PNM, by making 
a simulation with 66 neutrons in a periodic box. The EOS for nuclear 
matter as a function of the proton fraction $x_p=\rho_p/\rho$ is then 
parametrized as
 \begin{equation}
\label{eq:eos}
E(\rho,x_p)=E_{SNM}(\rho)+C_{s}\left(\frac{\rho}{\rho_0}\right)^{\gamma_{s}}
(1-2x_p)^2 \,.
\end{equation} 
 The two extra parameters of the symmetry energy term, $C_s$ and 
$\gamma_s$, were obtained by fitting $E(\rho,x_p=0)$ to the AFDMC result 
for PNM. This gives $C_s=31.3$ MeV and $\gamma_s=0.64$. Typical values for 
these parameters have been quoted as $C_s\approx31-33$ MeV and 
$\gamma_s\approx0.55-0.69$ by \cite{shetty07} and as $C_s=31.6$ MeV and 
$\gamma_s\approx0.69-1.05$ by \citet{worley08}. It should be noted that 
usually the symmetry energy is constrained over a range of densities 
typical of nuclei, whereas we have here fitted the parameters over a very 
wide density range. This means that the parametrization of eq. 
(\ref{eq:eos}) should be accurate up to very high densities.

In high-density matter, neutrons can produce protons and electrons by 
$\beta$ decay and so the equilibrium configuration can have a non-zero 
proton/neutron ratio, modifying the EOS away from that for PNM. The 
equilibrium concentration of protons $x_p$ can be computed by imposing
\begin{equation} 
\label{eq:betastab} 
\mu_n=\mu_p+\mu_e \,, 
\end{equation} 
 where $\mu_i$ is the chemical potential (for neutrons, protons and 
electrons respectively). For doing this, we consider the electrons as
comprising a relativistic Fermi-gas:
\begin{equation} 
\mu_e=[m_e^2+\hbar^2(3\pi\rho_e)^{2/3}]^{1/2} \,, 
\end{equation} 
and charge neutrality imposes that $\rho_e=x_p\rho$, where $\rho$ is the 
total nucleon density. The chemical potentials of the neutrons and protons 
are derived from equation (\ref{eq:eos}), and equation (\ref{eq:betastab}) 
is then solved to give $x_p$ as a function of $\rho$. Another 
consideration is that when $\mu_e$ becomes larger than the muon mass, the 
production of muons is favoured, and their contribution must then also be 
considered. For our EOS, muons begin to appear at $\rho\approx0.133$ fm$^{-3}$.

\begin{figure}
\vspace{0.5cm}
\begin{center}
\includegraphics[width=8cm]{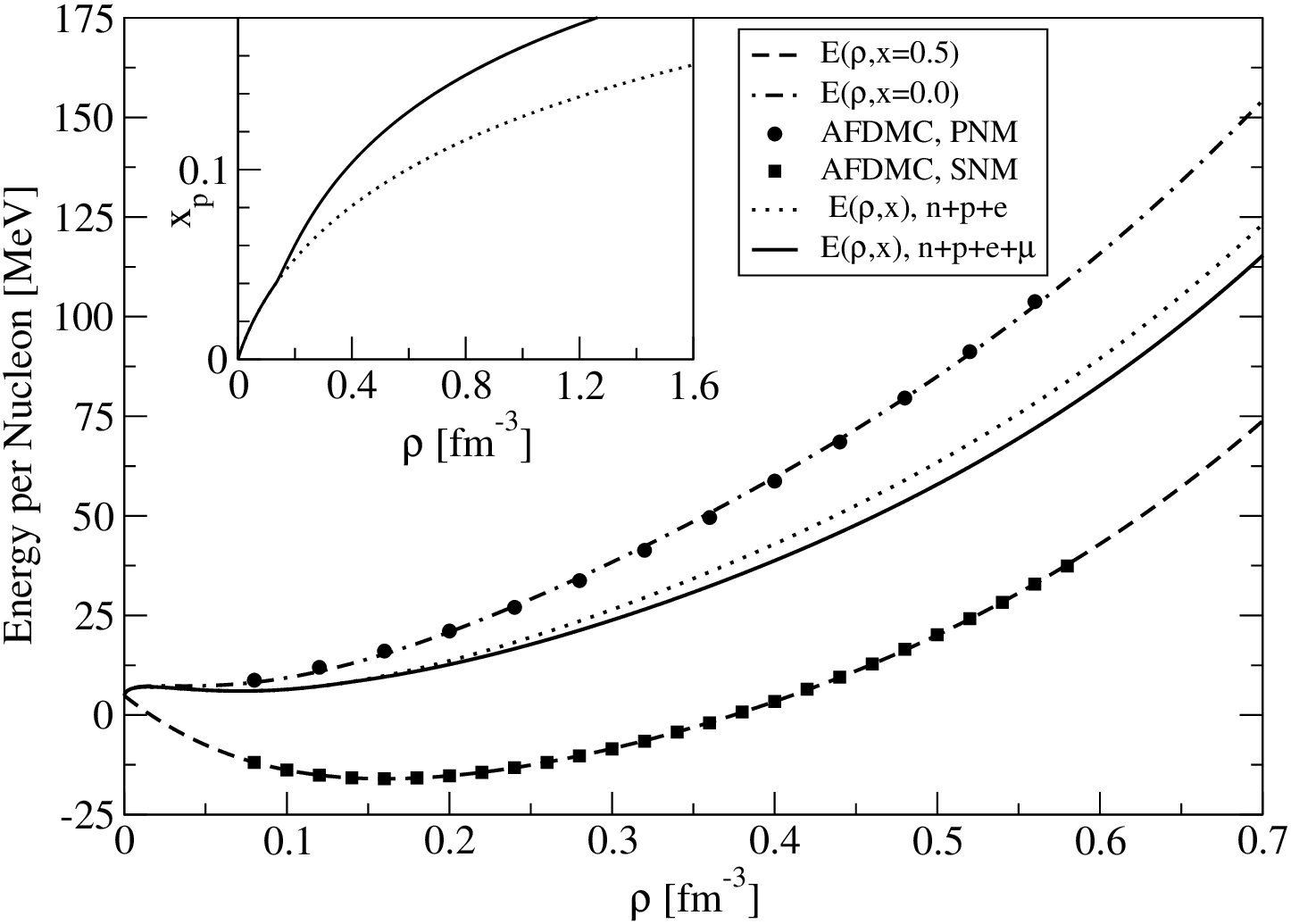}
\vspace{0.5cm}
\caption{The equation of state for symmetric nuclear matter (dashed 
line), pure neutron matter (dot-dashed line) and $\beta$-equilibrium 
nuclear matter with both electrons and muons (full line) and with 
electrons only (dotted line). The points show the AFDMC results for 
symmetric nuclear matter (SNM, squares) and pure neutron matter (PNM, 
circles) which have been used to fit the EOS. In the inset, we show the 
proton fraction $x_p$ plotted as a function of the total nucleon density, 
computed considering the presence of just electrons as negatively-charged 
particles (dotted line) and with both electrons and muons (full line).}
\label{fig:eos}
\end{center}
\end{figure}

The EOSs calculated for SNM, PNM and $\beta$-equilibrium matter are shown 
in Fig. \ref{fig:eos}, with the energy per nucleon being plotted against 
the nucleon number density $\rho$; in the inset, we show the proton 
fraction $x_p$ plotted as a function of $\rho$, computed considering both 
electrons and muons (full line) and only electrons (dotted line). The EOS 
for PNM is softer than that calculated with the AV8$'$+UIX potential by 
\citet{gandolfi09} using the same AFDMC many-body method and this then 
produces a similar behaviour for the $\beta$-equilibrium matter. The main 
reason for this comes from the different treatment of many-body effects.
 
At very high densities, the chemical potential of the nucleonic matter 
becomes larger than the threshold for creation of heavier particles. Such 
states are due to the up and down quarks transforming to strange quarks, 
so that particles with strangeness (hyperons) start to appear. A realistic 
EOS should include these when they appear and this can seriously modify 
the structure of the star. We do not include this in our present 
calculations (although we make some comment about its likely effects in 
the next section); we leave inclusion of this until a subsequent paper.

\section{Resulting neutron star models}
\begin{figure}
\vspace{0.5cm}
\begin{center}
\includegraphics[width=8cm]{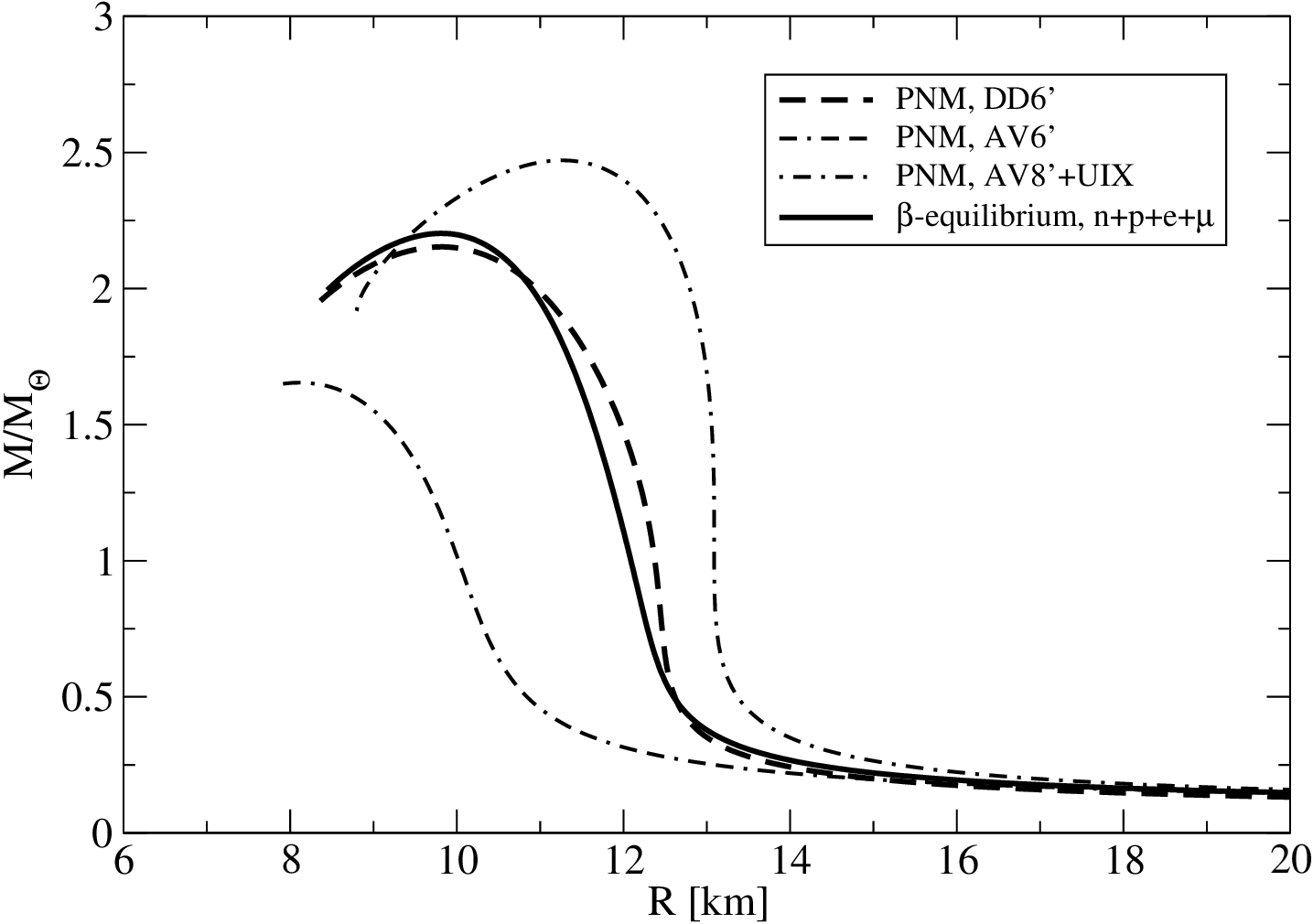}
\vspace{0.5cm}
 \caption{Predicted neutron-star masses (in units of M$_\odot$) 
plotted as a function of stellar radius (in km). Four different equations 
of state are considered: those discussed in this paper for 
$\beta$-equilibrium matter (full line) and pure neutron matter (PNM, 
DD6$'$, dashed line), the equivalent one for PNM with just two-body 
interactions (PNM, AV6$'$, dot-dash-dashed line), and a previous one from 
\citet{gandolfi09}, for PNM with three-body interactions (PNM, AV8$'$+UIX, 
dot-dashed line).}
 \label{fig:mvsr} 
\end{center} 
\end{figure}

 When the EOS of the neutron-star matter has been specified, the structure 
of an idealized spherically-symmetric neutron star model can be calculated 
by integrating the Tolman-Oppenheimer-Volkoff (TOV) equations:
\begin{equation}
\frac{dP}{dr}=-\frac{G[m(r)+4\pi r^3P/c^2][\epsilon+P/c^2]}{r[r-2Gm(r)/c^2]} \,,
\label{eq:tov1}
\end{equation}
\begin{equation}
\frac{dm(r)}{dr}=4\pi\epsilon r^2 \,,
\label{eq:tov2}
\end{equation}
where $P=\rho^2(\partial E/\partial\rho)$ and $\epsilon=\rho(E+m_N)$ are 
the pressure and the energy density, $m_N$ is the average nucleon mass, 
$m(r)$ is the gravitational mass enclosed within a radius $r$, and $G$ is 
the gravitational constant. The solution of the TOV equations for a given 
central density gives the profiles of $\rho$, $\epsilon$ and $P$ as 
functions of radius $r$, and also the total radius $R$ and mass $M=m(R)$. 
A sequence of models can be generated by specifying a succession of values 
for the central density. In Fig. \ref{fig:mvsr} we plot the mass $M$ 
(measured in solar masses M$_\odot$) as a function of the radius $R$ 
(measured in km), as obtained from calculations with four different 
prescriptions for the EOS: the $\beta$-equilibrium and PNM EOSs 
discussed in Sections 2 and 3, the equivalent one for PNM with just 
two-body interactions (using AV6$'$), and a previous one from 
\citet{gandolfi09}, for PNM with three-body interactions (using 
AV8$'$+UIX). Models to the right of the maximum of each curve are stable 
to radial perturbations and these are the ones of interest for 
astrophysical neutron stars. The maximum mass obtained with the 
two-body interaction AV8$'$ is very similar to that for AV6$'$.

It is interesting to make a comparison between these curves so as to see 
the changes caused by introduction of the various different features. The 
solid curve ($\beta$-equilibrium) is our best proposal for the 
neutron-star EOS but it can be seen that it differs only very little from 
the pure neutron matter EOS (where the radii for a given mass are just 
slightly larger within the main range of interest). There is a 
considerable difference, however, with respect to the previous AV8$'$+UIX 
curve for pure neutron matter, with the maximum mass being reduced from 
$\sim 2.5$ M$_\odot$ to $\sim 2.2$ M$_\odot$ and the radii in the main region 
of interest also being substantially reduced. This reflects the effective 
softening of the EOS caused by the different treatment of many-body 
effects.
 
An objective of this type of work is to attempt to constrain microphysical 
models for neutron-star matter by making comparison with astronomical 
observations \citep[see][]{lattimer07}. This is just starting to be 
possible now and further progress is anticipated within the next few 
years. At present, the only neutron stars for which masses are accurately 
known are the components of the best-observed binary pulsars, for which 
timing measurements give results correct to many significant figures. The 
maximum mass for any of these is the $1.441$ M$_\odot$ for the 
Hulse/Taylor binary pulsar PSR 1913+16 \citep[see][]{weisberg05}. However, 
there is accumulating evidence for higher masses, particularly for neutron 
stars in binary systems together with white dwarfs \citep[see][]{ransom05} 
and there is now a widespread belief that the maximum should probably be 
in the range $1.8$ M$_\odot - 2.1$ M$_\odot$ (at least when the rotation 
is sufficiently slow, as is the case for almost all pulsars so far 
observed). At high enough densities, it is expected that the composition 
of the matter would change because of the appearance of either hyperons or 
deconfined quarks, both of which are likely to decrease the maximum mass 
(see, for example, \citealp{schulze06} for the case of hyperons, and 
\citealp{akmal98} for an example of the inclusion of quark matter). The 
central density corresponding to our maximum mass for $\beta$-equilibrium 
matter is $\rho_c\approx1.2$ fm$^{-3}$ which is well within the range 
where these changes are likely to have occurred, and so we expect that the 
maximum mass would be slightly lower than that shown in Fig. 
\ref{fig:mvsr}. This brings us well within the expected range.

\begin{figure}
\vspace{0.5cm}
\begin{center}
\includegraphics[width=8cm]{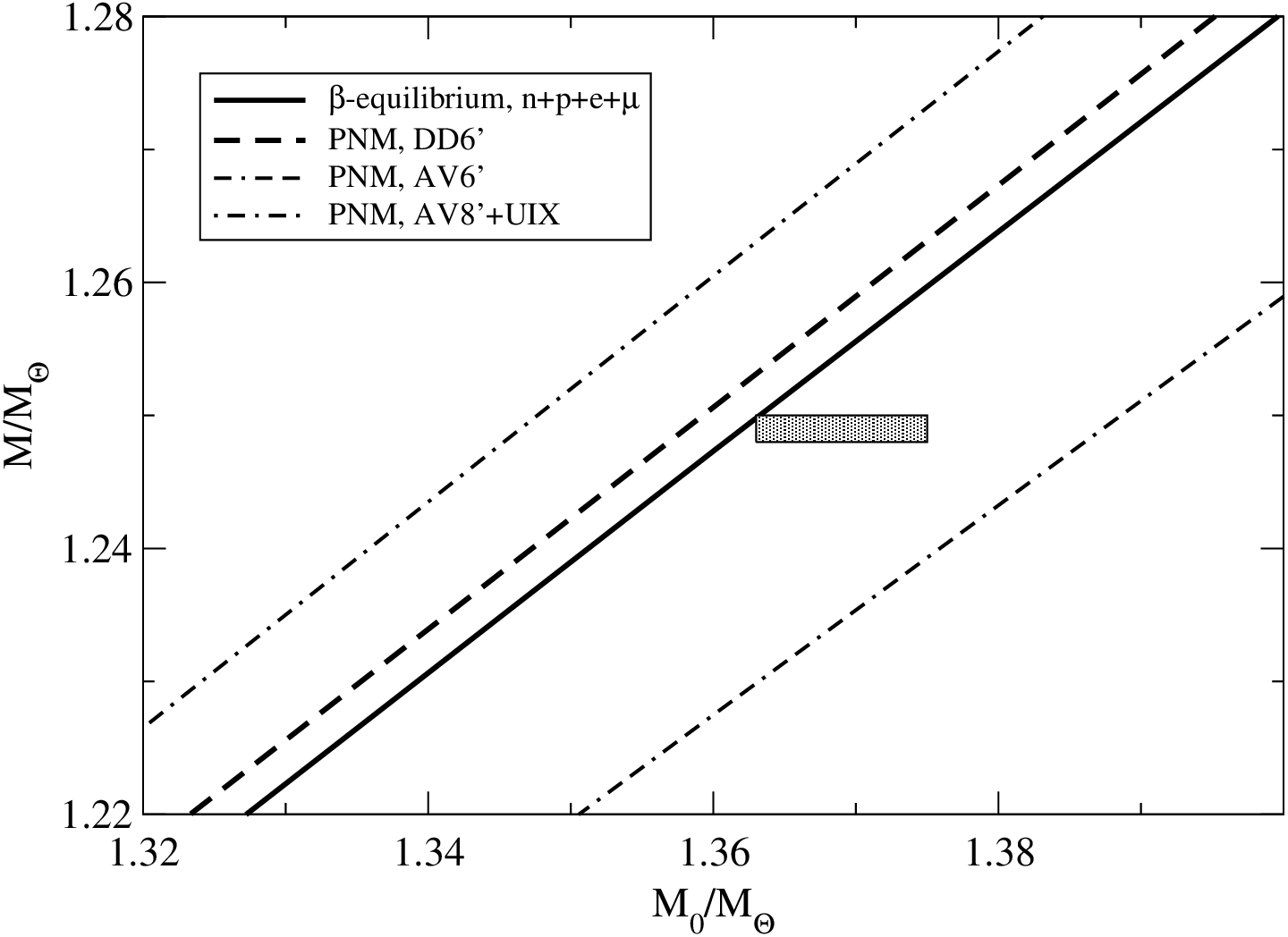}
\vspace{0.5cm}
 \caption{The gravitational mass of stellar models plotted as a 
function of the baryonic mass. See the caption of Fig. \ref{fig:mvsr} for 
details. The shaded rectangle is the error box inferred from observations 
of the lower-mass component of the double pulsar PSR J0737-3059 and 
associated modelling \citep[following][with the modification mentioned in 
the text]{podsi05}. \citet{kitaura09} have proposed a smaller error box 
just to the left of this, but we prefer the one shown here.}
 \label{fig:mvsm0} 
\end{center}
\end{figure}

Observational constraints for the radius are more problematic, but one of 
the best of these seems to be the indirect constraint suggested by 
\citet{podsi05} in the case of the less massive component of the double 
pulsar PSR J0737-3059. If, as seems likely, this neutron star was the 
product of an electron-capture supernova, then the total pre-collapse 
baryon number of the stellar core is rather precisely known from model 
calculations and, since only a very small loss of material is expected to 
have occurred in the subsequent collapse, the baryon number of the neutron 
star is itself also well-known. Together with the very accurate value for 
the gravitational mass (calculated from pulsar timing), this can be used 
to place a quite stringent constraint on the EOS. The baryon number $A$ of 
a neutron-star model can be readily calculated from 
 \begin{equation} 
\frac{{\rm d} A(r)}{{\rm d} r}=4 \pi \rho\,r^2\, \left(1-\frac{2 G m}{r 
c^2}\right)^{-\frac12}, \label{eq:tov3} 
\end{equation} 
 which needs to be solved together with equations (\ref{eq:tov1}) and 
(\ref{eq:tov2}) (we recall that $\rho$ is here the baryon number density). 
In practice, it is convenient to talk in terms of the {\emph baryonic 
mass}, defined as $M_0 = m_N A(R)$, rather than $A(R)$ itself: the 
difference between the baryonic mass and the gravitational mass depends on 
the compactness of the neutron star, and hence indirectly on the radius. 
If $M$ is plotted against $M_0$ for a given EOS, then the curve needs to 
pass through a certain error box in order to be consistent with the 
observations for PSR J0737-3059, subject to the assumptions being made in 
the analysis. This plot is shown in Fig. \ref{fig:mvsm0} for the same 
EOSs shown in Fig. \ref{fig:mvsr}. (We use here the error box as 
suggested by Podsiadlowski et al. 2005, extended to the left by $3 \times 
10^{-3}$ M$_\odot$ to include a plausible upper limit for the matter lost 
during the core collapse.) The curve for our $\beta$-equilibrium EOS just 
touches the top left-hand corner of the error box. However, the central 
density of our model corresponding to the mass of the pulsar concerned ($M 
= 1.249$ M$_\odot$) is more than three times nuclear matter density, by 
which point hyperons have probably already appeared, giving some softening 
of the EOS which would move the curve slightly downwards.

\section{Conclusions}
In this letter we have presented a new equation of state for neutron-star 
matter based on microscopic calculations made with the Auxiliary 
Field Diffusion Monte Carlo technique, using a semi-phenomenological 
Hamiltonian including a realistic two-body interaction and with the effect 
of many-body forces being included by means of additional 
density-dependent terms. We have presented results from stellar model 
calculations using the new EOS and some related variants, focusing on the 
mass/radius relation and the relation between the gravitational mass and 
the baryon number, and we have compared them with observational 
constraints.

\section*{Acknowledgments} 
 We thank S.~Reddy, R.~De Pietri and J.R.~Stone for useful discussions. 
This work was supported by CompStar, a Research Networking Programme of 
the European Science Foundation; KES thanks SISSA for kind hospitality and 
acknowledges support from NSF grant PHY-0757703. Calculations were 
performed using the HPC facility ``BEN'' at ECT$^\star$ in Trento, under a 
grant for Supercomputing Projects, and using the HPC facility of 
SISSA/Democritos in Trieste.

\bibliographystyle{mn2e}
\bibliography{nstar}
\bsp

\end{document}